# A Monte Carlo Simulation Study of Substrate Effect on AB Type Thin Film Growth

İzzet Paruğ Duru[*], Şahin Aktaş[2]

[*] Department of Physics, Marmara University, Turkey
parugduru@gmail.com
[2] Department of Physics, Marmara University, Turkey
saktas@marmara.edu.tr

**Abstract.** An iterative algorithm based on Monte Carlo method is used to model thin film growth of AB type molecule and crystallization. Primarly, PVD technique is investigated since it is one of the most preferred on thin film growth processes. The formation of thin film has been simulated for a cubic region with 10000 A type and 10000 B type atoms. Up to third nearest neighboring cells have been taken into account to realize the inter-atomic interactions. Boltzmann statistics is used to deal temperature effect by treating both A and B atoms as classical particles. The proposed substrate with the same crystal structure of the film was simulated by fixing the first layer of the film as having a perfect crystal structure. Roughness of the film surface is analyzed by sampling the RMS (Root mean square roughness) parameter both analytically and visually.

## 1 Introduction

Thin film of nanometer thickness has been developed by various coating techniques and has many applications in technology. Actual experimental study in nano scale requires expensive instrumentation. Alternatively thin film growth under various conditions can be studied even by using computational powers of today's personal computers. [1] In these studies, mostly Monte Carlo Method (MC) and Molecular Dynamics (MD) has been used to mimic the process. [2-4] MD has advantage of using energy and momentum conservation when inter-atomic interactions are handled. However this requires substantially high calculating power because the program must keep track of all particles energy and momentum all the time throughout the simulation process to conserve energy and momentum of the system. [5, 6] As a stochastic approach Monte Carlo method with Metropolis algorithm [7] can be more useful than a deterministic one (MD) for this study. Experimental studies give the end state of crystal structures without giving any clue about mid-states while ordered lattice structure evolves. On the other hand computational models enable us to trace film formation and crystallization process. [8]

Although actual physical interactions between atoms are very complicated, they can be approximated by symbolic parameters and then the whole process of formation of the thin film can be mimiced by Monte Carlo method. The interaction between the substrate atoms and the atoms of coating material studied by defining the crystall structure of substrate before the simulation starts. [2] Different type of potentials, Morse potential, Lennard-Jones potential and Yukawa potential as a function of the position of the atoms, have been used to sample interatomic forces while simulating the film gowth process. Effects of these potentials were discussed. [9] Generally the temperature effect handled with Boltzmann statistics in both Physical Vapor Deposition (PVD) and Chemical Vapour Deposition (CVD) simulations. [2, 3]

In MC method, only the small part (arbitrarily chosen) of the system is taken into account at each decision process. We use MC method with Metropolis algorithm to deal temperature effect (thermal fluctuations can be dealt better by using MC method). [7] We also study crystallization process by tracing the film growth through analyzing successive MC steps. We consider formation of thin film of two different type of atoms (AB type) by defining the bonding energy of different atoms is more attractive from the similar ones. In this study, thin film growth of chemical composition of two different atoms (AB) using PVD technique has been studied. The crystallization and filling ratios (coverage of z layers) have also been investigated to study surface roughness.

## 2 Method of Simulation

**2.1 PVD Lattice Model**

The simulated system is consisted of a cubic region of 50 unit length borders for each coordinate (x, y, z). 10000 A type, 10000 B type, totally 20000, particles are represented as A and B type atoms in the simulation region. Numerically, each particle has a label and all particle coordinate numbers are handled with matrices. At the beginning, particles are distributed randomly throughout the system. To start MC process a particle at random is chosen, then all possible reaction steps of the chosen atom are calculated, [10] up to third nearest neighboring cells through the moving direction are computed. By utilizing Metropolis algorithm a choice is made and the chosen atom is moved accordingly. When all particles has been processed, a Monte Carlo time step will be completed. The full simulation consists of a large number of such steps. The total physical time is expressed in the number of Monte Carlo steps. Gravitational effect is handled by taking tendency of moving downwards as 60% on vertical motion of particles. [11]

## 2.2 Interaction Potential Between Two Atoms

There are many dosens of deposition technologies about film formation. The *evaporative* ones handle thermal conditions for growth of thin film over the substrate material. We choosed PVD tecnique since it is one of the most preferred on the thin film growth processes. Temperature fluctuations were taken into account by using Boltzman statistics with MC method . [12-14]

$$R_p = \frac{e^{-E_a/kT}}{\sum_{E_a} e^{-E_a/kT}} \quad (1)$$

The binding probability, $R_p$, is proportional to temperature (T) given by Boltzmann factor. $E_a$, denotes binding energy between the particles.

In our simulation method, the most important part is to mimic interatomic forces which would ultimately determine binding energy $E_a$ with suitable models to write a feasible algorithm considering limited calculation capacity of desktop computers. [3] Since we study the film of chemical composition of AB type, we must somehow distinguish the two cases of having interactions between similar and dissimilar atoms. AB type of atomic combination means that binding energy between atom A and atom B must be stronger than the attraction between similar atoms (A-A or B-B). [11]

| 2 | 3 | 4 | 5 |
|---|---|---|---|
| 1 | S(i,j,k) 0 | S(l,m,n) 11 | 6 |
| 10 | 9 | 8 | 7 |

**Fig. 1.** Demonstration of movement decision of a particle for the first neighboring cells

When the atom located in the cell represented by parameters (i, j, k) is chosen. The cells labelled as '2', '5', '7', '10' have not been taken into account while computing the probability of the movement of the particle that exists at "0" labelled cell to "11". Although movement of a particle can be in any dimension in cubic lattice cell, an xy plate shown in Fig. 1 is used to demonstrate the event. All of the possible motion directions of the chosen atom have been taken into account and a decision is made by using Metropolis algorithm. $E_0^1$ represents interaction chain of atoms located in cells of '0' and '1'.

$$\Delta E_{AB} = E_0^1 + E_0^3 + E_0^9 + E_0^{-z} + E_0^{+z} - [E_{11}^4 + E_{11}^6 + E_{11}^8 + E_{11}^{-z} + E_{11}^{+z}] \qquad (2)$$

When $\Delta E_{AB} < 0$ condition is satisfied, a movement to the cite '11' will be made. While determining all these threshold energy differences, we assumed that the combination probability of the similar atoms is less than the different ones which is mentioned on the beginning of this section. $E_0^{-z}, E_0^{+z}, E_{11}^{-z}, E_{11}^{+z}$ are the threshold energies of the particles at the z layer neighbour of '0' labelled cell.

### 2.3 Effect of Substrate

Suitable substrate is important in point of the good quality thin film growth. Similar crystallographic orientation that exists between substrate atoms and the source materials improves the crystallization ratio and filling ratio (coverage of z layers). [12] [15] Hence, we have integrated a suitable substrate to our model to investigate the effect of substrate on crystallization.

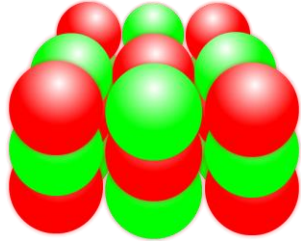

**Fig. 2.** 3D view of suitable substrate

The suitable substrate is formed with the well-ordered particles on the first layer. The ordered positions of atoms in this layer is kept fixed during simulation. A homoepitaxial process modelled for the film growth over the substrate material. [11]

### 2.4 Measurement of Crystallization

Crystallization has been determined by studying the type of the neighbour of filled cell. There are 6 nearest-neighbour in 3D dimensional cubic symmetry. If all 6 neighbouring cells contain opposite type of atoms, we assign '+1' to CP (crystallization parameter). We supposed that this condition pictures the perfect crystallization for an atom.

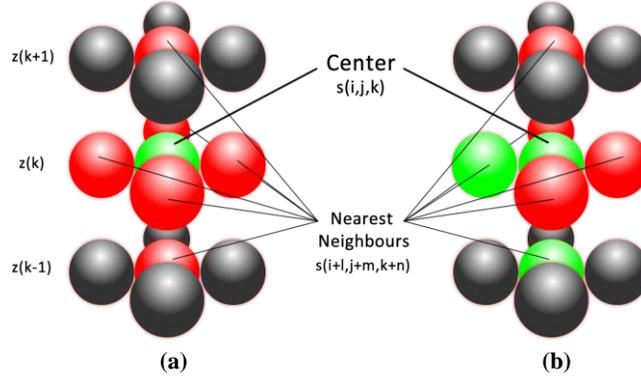

**Fig. 3. (a)** Type of perfect crystallization: 6 dissimilar atoms (B type) **(b)** partial crystallization: 4 dissimilar atoms (B type) and 2 similar (A type) atoms that neighbouring to S(i,j,k) (A type) Gray ones haven't been taken into account.

The number of dissimilar atoms that exist at neighbouring cells divided by total neighbour numbers and summed up with CP. The percentage ratio of CP for a certain level can be obtained as,

$$\overline{CP} = 100 \times \frac{1}{N}\sum_{i=1}^{N} CP_i, \qquad N=2500 \qquad (3)$$

Equation (3) mentions about the crystallization for a 'z' layer in 3D coordinate space. [11] The most suitable number of MC steps was determined as 2000 MC steps by trial and error method to optimize computational time with the stability of the end results. [14]

**2.5 Measurement of Surface Roughness**

Surface roughness can be used to estimate the quality of thin films. [16] During the film growth process the temperature effect against the crystallization and the morphology of film can be studied by analyzing the surface roughness of film with any z range height. [17] [18] Surface roughness parameters of the film structure are calculated by using *'connected component'* technology in MATLAB medium. Firstly, we looked for the occupation state of the particles at nearest neighboring cells. The particles which hadn't any connection with the crystal bulk haven't been taken into account while calculating surface roughness of film. For calculation of the RMS (root mean square roughness) parameter of the film, first μ is calculated as

$$\mu = \sqrt{\frac{1}{N_x N_y}\sum_{i=1}^{N_x}\sum_{j=1}^{N_y} z(x_i, y_j)} \qquad N_x = 50 \qquad N_y = 50 \qquad (4)$$

$N_x$ and $N_y$ are the boundary values of the x and y coordinates respectively. μ also indicates the average thickness of the film on the surface region. z(x, y) represents amplitude of the cell that includes any type of atoms either A or B. Thus RMS of the film surface can be calculated by standard deviation of z(x, y).

$$S_q = \sqrt{\frac{1}{N_x N_y} \sum_{i=1}^{N_x} \sum_{j=1}^{N_y} (z(x_i, y_j) - \mu)^2} \qquad (5)$$

$S_q$, is the RMS value of the film surface settled between maximum and minimum peaks of the atoms existed on the upper layer of the film. Film surface roughness can be tuned by the film formation parameters relatively.

## 3 Results

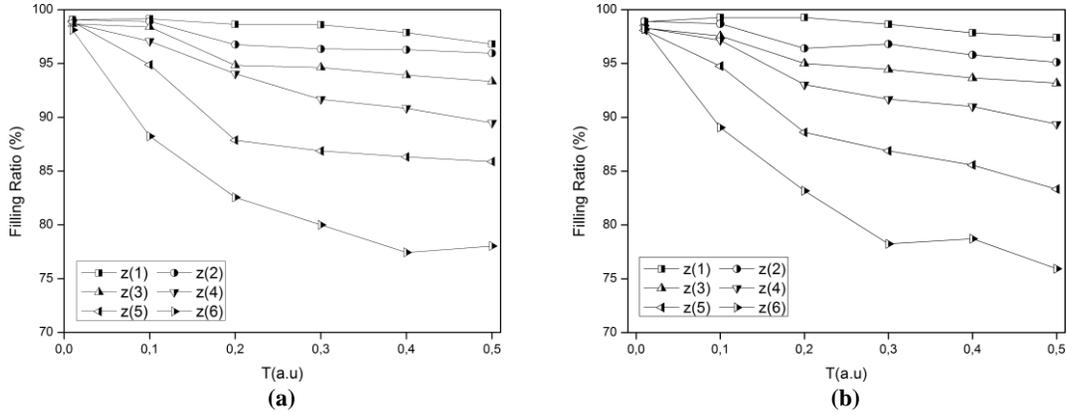

**Fig. 4.** Filling ratio of the first 6 layer according to the temperature scale **(a)** without suitable substrate **(b)** with suitable substrate

We studied temperature effect on crystallization. At low temperatures close to zero Kelvin, the filling factor was not affected much as seen Fig. 4. Nevertless, crystallization shows a different behaviour than filling factor as seen Fig. 5, which gives the impression that the crystallization (long-range order) has a discontinuity at a certain temperature that may be interpreted as a phase transition on crystallization.

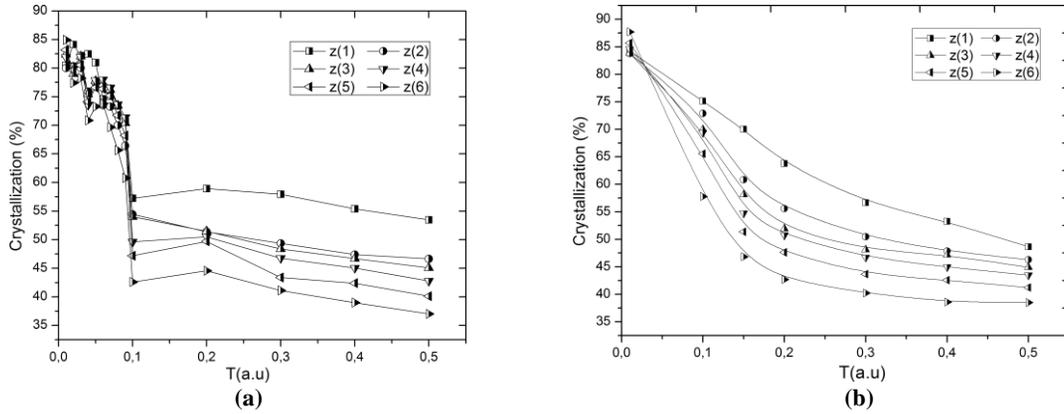

**Fig. 5.** Crystallization ratio of the first 6 layer according to the temperature scale **(a)** without suitable substrate **(b)** with suitable substrate

According to the Fig. 5, T=0.1 a.u is attributed as a critical value that mention about a phase transition of crystallization. Filling ratio given by Fig. 4 and crystallization given by Fig. 5 suggest that using a suitable substrate greatly improves crystallization of film forming on a surface compared with the care of without a special substrate used. Meanwhile the crystallization ratio that's scaled with T=0.01 a.u have the best peak as seen in Fig. 5. The first 8 layer of z dimension analysed with the T=0.01 a.u and T=0.15 a.u values more than the others due to their significant effects.

Fig. 6 shows the actual positions of atoms on a given layer with and without suitable substrate. The effect can be even detected visually.

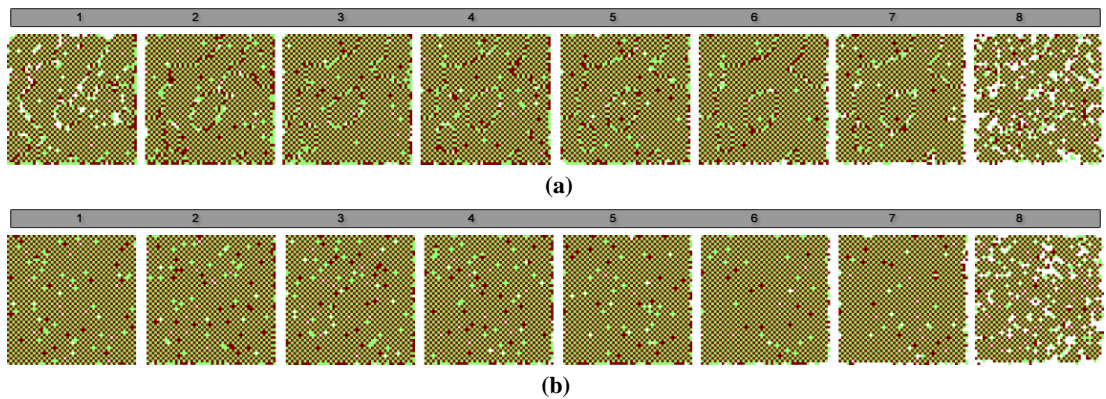

**Fig. 6.** Particle distribution of first 8 layer of thin film at T=0.01 a.u temperature **(a)** without suitable substrate **(b)** with suitable substrate

Fig. 6b shows that at low temperature, using suitable substrate results with a better crystallization of layers. Homoepitaxial process have an advantage to redound the crystallization of z layers more than the other techniques.

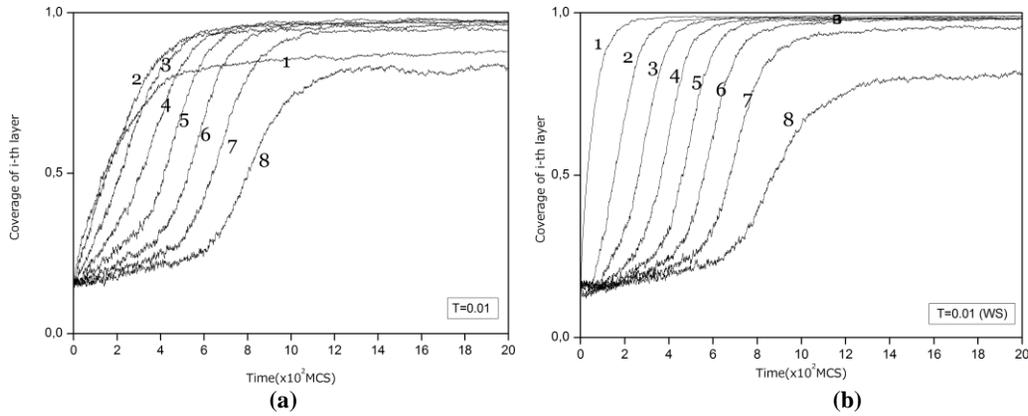

**Fig. 7.** Coverage of the z layers against the MC step **(a)** without suitable substrate **(b)** with suitable substrate at T=0.01 temperature.

We investigated an efficient Monte Carlo Step (MCS) number to decrease computing time. 2000 is MCS is enough for our system to reach equilibrium. Suitable substrate fastens the equilibrium process as seen in Fig. 7(a-b) and Fig. 6(a-b).

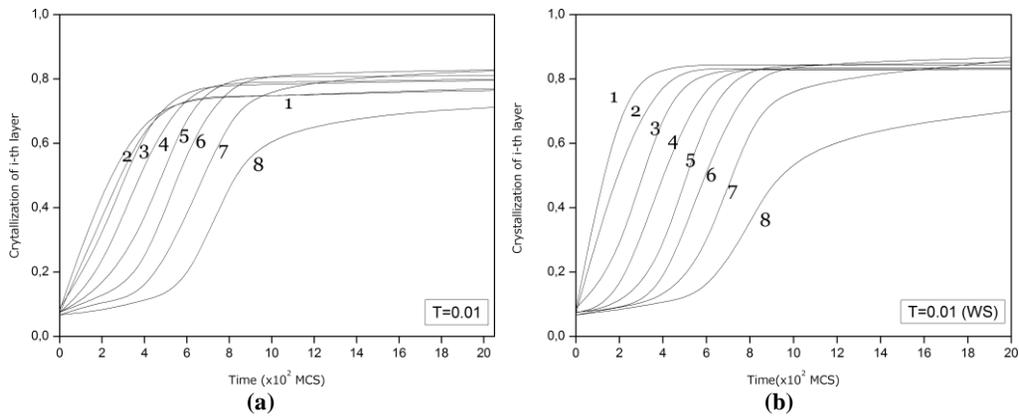

**Fig. 8.** Crystallization ratio of the z layers against the MC step **(a)** without suitable substrate **(b)** with suitable substrate at T=0.01 a.u temperature.

It's clear that using substrate greatly improves crystallization that can even be detected visually. Fig. 9 compares crystallization ratio of two cases visually that clearly shows the improving role of using proper substrate.

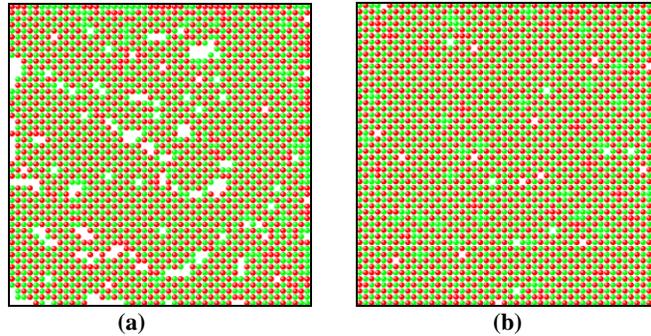

**(a)**            **(b)**

**Fig. 9.** Particle distribution of first layer at T=0.01 a.u temperature **(a)** without suitable substrate **(b)** with suitable substrate.

The lattice defaults during the crsytallization process above substrate plane minimized with homoepitaxy as seen on Fig. 9(a-b). [11] Secondly effect of the suitable substrate to the surface roughness of the film can be easily seen by comparing Fig. 10a and Fig. 10b. Two surface can be compared by using color bar. The light tones of the color bar is used to point out the peaks on +z direction. Through the bottom of the color bar the negative peaks can be seen as holes.

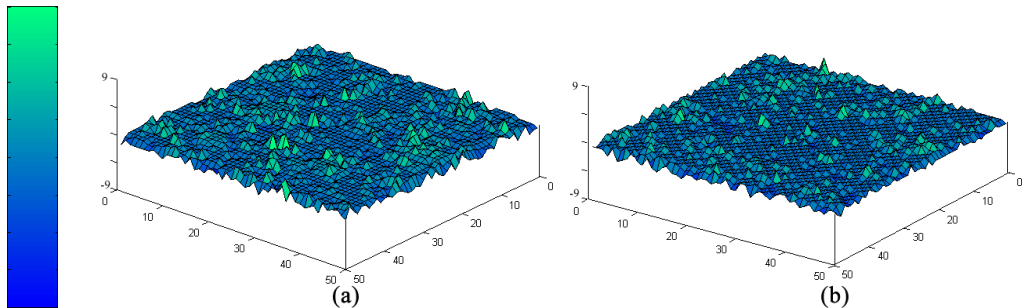

**(a)**            **(b)**

**Fig. 10.** Surface roughness graphic at T=0.01 a.u temperature **(a)** without suitable substrate **(b)** with suitable substrate

With a suitable substrate RMS value is calculated as 0.009 a.u, otherwise it's 0.043 a.u. At higher temperature the long-range order for both with/without substrate has been lost at this range of temperature as seen in Fig. 11(a-b).

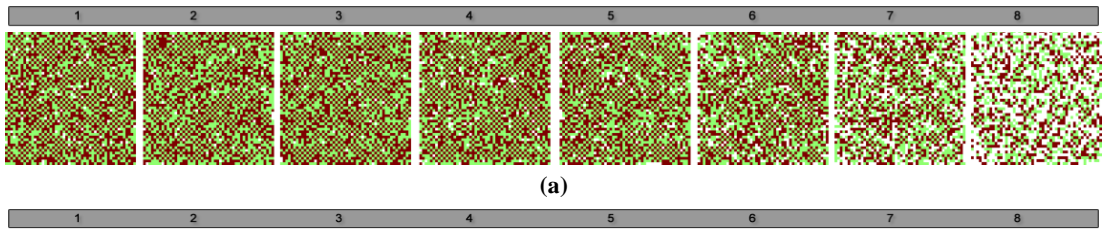

**(a)**

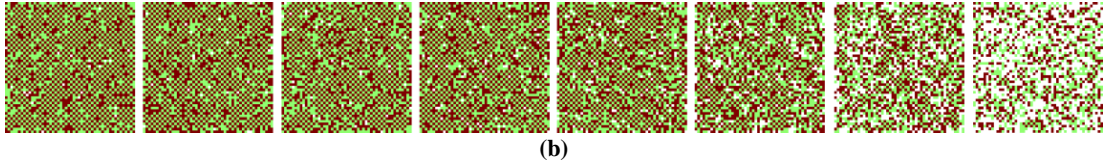
**(b)**

**Fig. 11.** Particle distribution of first 8 layer of thin film at T=0.15 a.u temperature **(a)** without suitable substrate **(b)** with suitable substrate

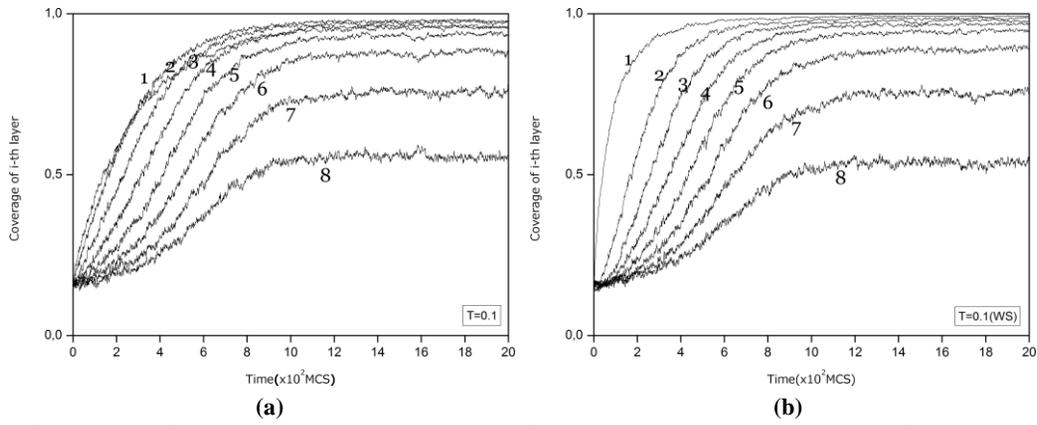

**(a)**                          **(b)**

**Fig. 12.** Coverage of the z layers against the MC step for first layer **(a)** without suitable substrate **(b)** with suitable substrate at T=0.15 a.u temperature.

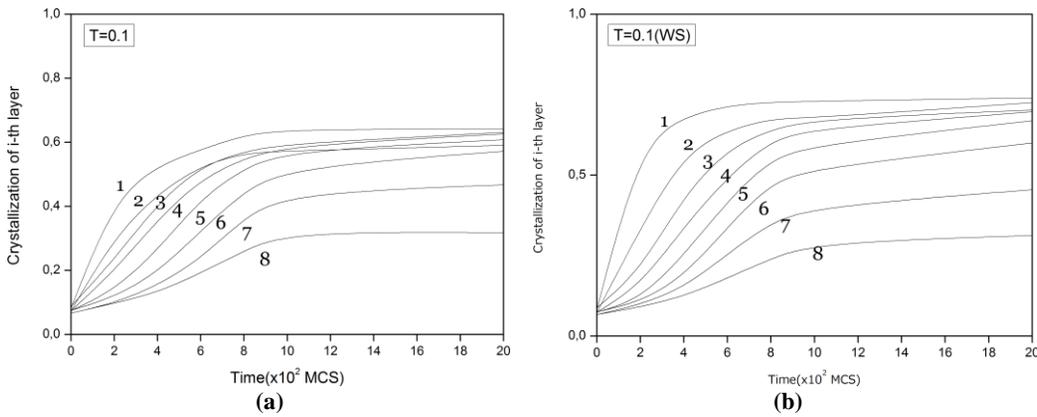

**(a)**                          **(b)**

**Fig. 13.** Crystallization ratio against the MC step for first layer **(a)** without suitable substrate **(b)** with substrate at T=0.15 a.u temperature

According to Fig. 12a and Fig. 12b, first layer of the film reached to equilibrium priorly. It can be easily understood from Fig. 14 visually that the crystallization ratio has a bigger value when a suitable substrate is preferred.

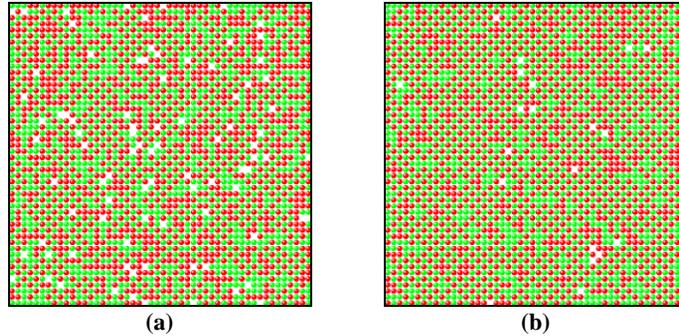

**Fig. 14.** Particle distribution of first layer (z=1) at T=0.15 a.u **(a)** without suitable substrate **(b)** with suitable substrate (WS)

The surface roughness of the film deposited at T=0.15 a.u temperature analysed. From Fig. 15a and Fig. 15b even though the suitable substrate smoothens the surface, at high temperatures roughness of the film increases in both cases. With a suitable substrate RMS value is calculated as 0.0432, otherwise it's 0.0047 at T=0.15 a.u.

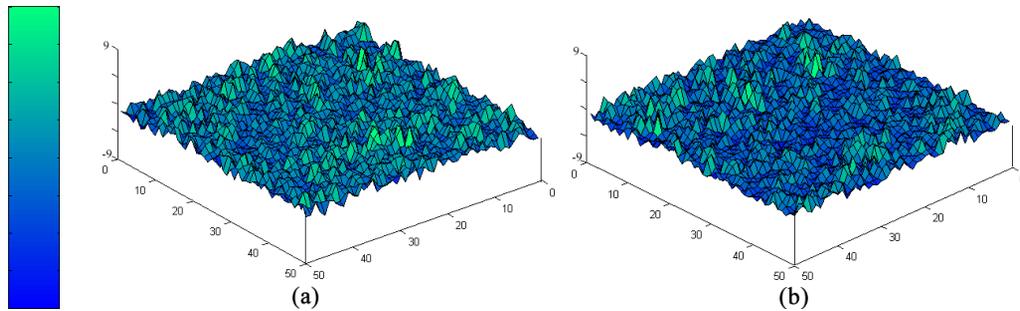

**Fig. 15.** Surface roughness graphic at T=0.15 a.u temperature **(a)** without suitable substrate **(b)** suitable substrate

## 4 Conclusion

We have studied cubic lattice symmetry by using Monte Carlo method to study the crystallization and surface roughness of a thin film composed of two different types of atoms (AB type molecule) grown by any PVD technique epitaxially. Our calculations showed that the system has evolved after 2000 MC step past, and particles have formed a rock-salt like crystal structure at low temperatures. As a critical value T=0.15 a.u showed a phase transition about crystallization of film. We found that using a suitable substrate increased the crystallization and filling ratio of AB type molecules at all z layers. In this respect the surface of the films which deposited on a suitable substrate is less rough than using an amorphous or random substrate.